\documentclass[english]{article}
\usepackage[T1]{fontenc}
\usepackage[latin9]{inputenc}
\usepackage{textcomp}
\usepackage{mathrsfs}
\usepackage{amsmath}
\usepackage{amssymb}

\makeatletter

\DeclareRobustCommand{\greektext}{%
  \fontencoding{LGR}\selectfont\def\encodingdefault{LGR}}
\DeclareRobustCommand{\textgreek}[1]{\leavevmode{\greektext #1}}
\DeclareFontEncoding{LGR}{}{}
\DeclareTextSymbol{\~}{LGR}{126}

\newcommand{\lyxaddress}[1]{
\par {\raggedright #1
\vspace{1.4em}
\noindent\par}
}

\makeatother

\usepackage{babel}
\begin{document}

\title{Solutions of the Schrödinger equation, boundary condition at the
origin, and theory of distributions.}

\author{Y. C. Cantelaube%
\thanks{e-mail : yves.cantelaube@univ-paris-diderot.fr%
}}

\maketitle

\lyxaddress{U.F.R. de Physique, Université Paris Diderot, Bâtiment Condorcet,
75205 Paris cedex 13, France}
\begin{abstract}
In a central potential the usual resolution of the Schrödinger equation
in spherical coordinates, whose solutions are of the form \textgreek{Y}
= \textit{R}(\textit{r})$\mathit{Y_{l}^{\mu}\left(\theta,\varphi\right)}$
= {[}\textit{u}(\textit{r})/\textit{r}{]}$\mathit{Y_{l}^{\mu}\left(\theta,\varphi\right)}$,
consists in substituting in $\Psi$ the solutions \textit{R}(\textit{r})
or \textit{u}(\textit{r}) of the radial equations considered as the
radial parts of the Schrödinger equation. However, the solutions must
be supplemented with the boundary condition \textit{u}(0) = 0 in order
to rule out singular solutions. There is still no consensus to justify
this condition, with good reason. It is based on a misunderstanding
that comes from the fact that the radial equation in terms of \textit{R}(\textit{r})
is derived from the Schrödinger equation, and the radial equation
in terms of \textit{u}(\textit{r}) from the former, by taking the
Laplacians of $\Psi$ and of\textit{ R}(\textit{r}) in the sense of
the functions. By taking these Laplacians in the sense of the distributions,
as it is required, we show that the radial equations are derived from
the Schrödinger equation when their solutions are regular, but not
when they are singular, so that the equations need not be supplemented
with any supplementary condition such as \textit{u}(0) = 0.
\end{abstract}

\section{Introduction : The usual resolution of the SE in spherical coordinates.}

In a central potential \textit{V}(\textit{r}), to which we shall confine
ourselves, and which will be denoted by $\mathit{V_{r}}$, the operators
\textit{H}, \textbf{L}\texttwosuperior{} and\textit{ $L_{z}$ }all
commute with each other, so that the eigenfunctions of \textit{H},
or the solutions of the Schrödinger equation (SE), are of the separable
form

\begin{equation}
\Psi(r,\theta,\varphi)=\mathit{R\left(r\right)\mathit{Y_{\ell}^{\mu}\left(\theta,\varphi\right)}}=\frac{u(r)}{r}\mathit{Y_{\ell}^{\mu}\left(\theta,\varphi\right)}
\end{equation}

where the factor 1/\textit{r} is introduced for convenience, and the
spherical harmonics $\mathit{Y_{l}^{\mu}\left(\theta,\varphi\right)}$
are eigenfunctions of the square of the angular momentum

\begin{equation}
\mathbf{L}^{2}\mathit{Y_{\ell}^{\mu}\left(\theta,\varphi\right)}=\ell(\ell+1)\hbar^{2}\mathit{Y_{\ell}^{\mu}\left(\theta,\varphi\right)}
\end{equation}

The usual resolution of the SE in spherical coordinates,

\begin{equation}
H\Psi=\left[-\frac{\hbar^{2}}{2m}\frac{1}{r}\frac{\partial^{2}}{\partial r^{2}}r+\frac{\mathbf{L}^{2}}{2mr^{2}}+\mathit{V}_{r}\right]R\left(r\right)\mathit{Y_{\ell}^{\mu}\left(\theta,\varphi\right)}=E\, R\left(r\right)\mathit{Y_{\ell}^{\mu}\left(\theta,\varphi\right)}
\end{equation}

is based on \textit{two assumptions.}

The \textit{first} assumption is that in any case the radial and angular
parts of the Laplace operator \textit{separately} act, and hence that
one \textit{may substitute} (2) in the expression of the Laplacian
of $\Psi$ which then amounts to a radial part times $\mathit{Y_{l}^{\mu}\left(\theta,\varphi\right)}$,
that is,

\begin{equation}
\Delta\Psi=\left[\frac{1}{r}\frac{\partial^{2}}{\partial r^{2}}r-\frac{\mathbf{L}^{2}}{\hbar^{2}r^{2}}\right]R\left(r\right)\mathit{Y_{\ell}^{\mu}\left(\theta,\varphi\right)}=\left[\frac{1}{r}\frac{d^{2}}{dr^{2}}r-\frac{\ell(\ell+1)}{r^{2}}\right]\mathit{R\left(r\right)\mathit{Y_{\ell}^{\mu}\left(\theta,\varphi\right)}}
\end{equation}

as also the Hamiltonian of $\Psi$,

\[
H\Psi=\left[-\frac{\hbar^{2}}{2m}\frac{1}{r}\frac{\partial^{2}}{\partial r^{2}}r+\frac{\mathbf{L}^{2}}{2mr^{2}}+\mathit{V}_{r}\right]R\left(r\right)\mathit{Y_{\ell}^{\mu}\left(\theta,\varphi\right)}
\]

\begin{equation}
=\left[-\frac{\hbar^{2}}{2m}\frac{1}{r}\frac{d^{2}}{dr^{2}}r+\frac{\ell(\ell+1)\hbar^{2}}{2mr^{2}}+\mathit{V}_{r}\right]R\left(r\right)\mathit{Y_{\ell}^{\mu}\left(\theta,\varphi\right)}
\end{equation}

The SE then reads

\begin{equation}
H\Psi=\left[-\frac{\hbar^{2}}{2m}\frac{1}{r}\frac{d^{2}}{dr^{2}}r+\frac{\ell(\ell+1)\hbar^{2}}{2mr^{2}}+\mathit{V}_{r}\right]R\left(r\right)Y_{\ell}^{\mu}\left(\theta,\varphi\right)=E\, R\left(r\right)\mathit{Y_{\ell}^{\mu}\left(\theta,\varphi\right)}
\end{equation}

which gives the radial equation

\begin{equation}
\left[-\frac{\hbar^{2}}{2m}\frac{1}{r}\frac{d^{2}}{dr^{2}}r+\frac{\ell(\ell+1)\hbar^{2}}{2mr^{2}}+\mathit{V}_{r}\right]R\left(r\right)=\mathit{E}\, R\left(r\right)
\end{equation}

The \textit{second }assumption is that by substituting \textit{u}(\textit{r})
and multiplying both sides of (7) by \textit{r}, the latter amounts
to the radial equation

\begin{equation}
\left[-\frac{\hbar^{2}}{2m}\frac{d^{2}}{dr^{2}}+\frac{\ell(\ell+1)\hbar^{2}}{2mr^{2}}+\mathit{V}_{r}\right]u\left(r\right)=E\,\mathit{u\left(r\right)}
\end{equation}

The resolution of the SE (3) in the three variables $\mathit{r}$,$\theta$,$\varphi$,
is thus replaced by the resolution of (7) or (8) , considered as the
radial parts of the SE, or the radial SE, depending only on the variable
\textit{r}. The solutions, substituted in $\Psi$, are written in
the form of series, $\mathit{R\left(r\right)}$= $r^{s}{\textstyle \Sigma_{k\geq0}a_{k}r^{k}}$,
and $\mathit{u\left(r\right)}$=$r{}^{s+1}$$\Sigma_{k\geq0}a_{k}r^{k}$,
by considering that for potentials which are not more singular at
the origin than 1/\textit{r}, the indicial equation gives the two
roots \textit{s }= $\mathit{\ell}$ and\textit{ s} = - ($\ell$+1).
For $\ell$ > 0 only the solutions given by the first root are normalizable.
In the radial case ($\ell$ = 0) $Y_{0}^{0}=$1/$\sqrt{4\pi}$, the
equation (8), which amounts to

\begin{equation}
\left[-\frac{\hbar^{2}}{2m}\frac{d^{2}}{dr^{2}}+\mathit{V}_{r}\right]u\left(r\right)=\mathit{E\,}u\left(r\right)
\end{equation}
is called the ``reduced Schrödinger equation''. Its normalizable
solutions are given by both roots, but for \textit{s }= -1, as \textit{u}(0)
= $\mathit{a_{o}}$ $\neq$ 0, \textit{R}(\textit{r}) and $\Psi$
behave at the origin like 1/\textit{r}, the solutions are ruled out
as inadmissible, so that the equations are supplemented with the boundary
condition

\begin{equation}
u(0)=0
\end{equation}

The question has always been to \textit{justify this supplementary
condition}. If these functions $\Psi$ are considered indeed as solutions
of the SE , the question is to justify that normalizable solutions
of the SE are not admissible, and if these functions are not considered
as solutions of the SE, the question is to justify that \textit{u}(\textit{r})
is nevertheless a solution of the reduced SE. This is why various
arguments, often conflicting, have been proposed to justify this condition,
but no one is satisfactory. It comes from the fact that the usual
assumptions, and hence the equations which are based on them, hold
in R$^{3}$/\{0\}, but \textit{not necessarily in R$^{3}$}.

\section{The Hamiltonian in R$^{3}$.}

Let us briefly recall that if $\Phi\in$ C$^{\infty}$(R$^{3}$) and
\textit{s }is a negative real, $\Psi$= $r^{s}\Phi$, which is of
C$^{\infty}$(R$^{3}$/\{0\}), but not of C$^{\infty}$(R$^{3}$),
defines in R$^{3}$ a distribution called ``pseudofunction''{[}1,2{]},
denoted by \textit{Pf.}$\Psi$, and its Laplacian in R$^{3}$ is the
Laplacian in the sense of the distributions. In the case $\Phi$=
1 the theory of distributions shows that depending on whether or not
\textit{s }is an odd negative integer, its Laplacian has a different
expression{[}1{]}, which can be written{[}3{]}

\[
\Delta Pf.r^{s}=s(s+1)Pf.r^{s-2}+\chi_{p}C_{p}\Delta^{p}\delta\qquad C_{p}=-\frac{(4p+1)\pi^{3/2}}{2^{2p-1}p!\,\Gamma(p+3/2)}
\]

\begin{equation}
\chi_{p}=\begin{cases}
1\quad & if\qquad p\in N\\
0 & if\qquad p\notin N
\end{cases}\qquad\qquad p=-\frac{s+1}{2}
\end{equation}

where $\Delta^{p}$ is the iterated Laplacian, and where the symbol
$\mathit{Pf}$. is used because depending on the value of \textit{s
}we have an ordinary function $\mathit{r^{s}}$, or a pseudofunction
$\mathit{Pf}$.$\mathit{r^{s}}$. The significance of the term \textit{s}(\textit{s}+1)\textit{Pf.}$r{}^{s-2}$
is the following. Just as $\mathit{r^{s}}$, which is of C$^{\infty}$(R$^{3}$/\{0\}),
but not of C$^{\infty}$(R$^{3}$) if \textit{s} < 0, defines in R$^{3}$
the pseudofunction \textit{Pf.}$\mathit{r^{s}}$, its Laplacian in
R$^{3}$/\{0\}, which is the Laplacian in the sense of the functions,
$\Delta r^{s}=s(s+1)r^{s-2}$, defines in R$^{3}$ the pseudofunction
\textit{Pf.}$\Delta r^{s}=$\textit{ s}(\textit{s}+1)\textit{Pf.}$r{}^{s-2}$.
Eq.(11) therefore also reads

\begin{equation}
\Delta Pf.r^{s}=Pf.\Delta r^{s}+\chi_{p}C_{p}\Delta^{p}\delta\qquad\qquad p=-\frac{s+1}{2}
\end{equation}

The symbol \textit{Pf.} can be dropped in the case of pseudofunctions
less singular at the origin than 1/$r{}^{3}${[}1{]}, but in the case
of the operator \textit{Pf}.\textgreek{D} only if it is equivalent
to \textgreek{D}, e.g. when it acts on a function of C$^{\infty}$(R$^{3}$).
Thus, \textit{Pf.}(1/\textit{r}) reads as\textit{ }1/\textit{r}, but\textit{
Pf.}\textgreek{D}(1/\textit{r}) is different from \textgreek{D}(1/\textit{r}),
for \textit{s }= -1, (12) gives

\begin{equation}
\Delta(\frac{1}{r})=Pf.\Delta(\frac{1}{r})+C_{o}\delta=-4\pi\delta\qquad\qquad Pf.\Delta(\frac{1}{r})=0
\end{equation}

The pseudofunction

\begin{equation}
Pf.R\left(r\right)=Pf.r^{s}\sum_{k\geq0}a_{k}{\displaystyle {\textstyle r^{k}}}
\end{equation}

where \textit{s} is any real, is written

\[
Pf.R\left(r\right)=Pf.[r^{s}S_{e}\left(r\right)]+Pf.[r^{s+1}S_{d}\left(r\right)]
\]

where $\mathit{S_{e}\left(r\right)={\textstyle \sum_{k\geq0}a_{2k}r^{2k}}}$and
$\mathit{S_{d}\left(r\right)=\sum_{k\geq0}a_{2k+1}r^{2k}}$ are of
C$^{\infty}$(R$^{3}$), so that $\Delta Pf.r^{s}S_{e}\left(r\right)$
and $\Delta Pf.r^{s+1}S_{d}\left(r\right)$, which are the Laplacians
of the product of a distribution by a function of C$^{\infty}$(R$^{3}$),
are given by the usual expression of the Laplacian of the product
of two functions. It enables to show that{[}3{]}

\begin{equation}
\Delta Pf.R\left(r\right)-Pf.\Delta R\left(r\right)=Q_{s}\left(\delta\right)
\end{equation}

where
\[
Q_{s}\left(\delta\right)=\chi_{p}\left[(C_{p}+2L_{p}r\frac{d}{dr})S_{e}\left(r\right)\right]\Delta^{p}\delta+\chi_{q}\left[(C_{q}+2L_{q}r\frac{d}{dr})S_{_{d}}\left(r\right)\right]\Delta^{q}\delta
\]

\begin{equation}
L_{p}=\frac{C_{p-1}}{8p(2p+1)}-\frac{C_{p}}{4}\qquad C_{-1}=0\qquad p=-\frac{s+1}{2}\qquad q=-\frac{s+2}{2}
\end{equation}

Consider now the pseudofunction

\begin{equation}
Pf.\Psi=Pf.R\left(r\right)\mathit{Y_{\ell}^{\mu}\left(\theta,\varphi\right)}{\displaystyle {\textstyle =Pf.r^{s}\sum_{k\geq0}a_{k}r^{k}}\mathit{Y_{\ell}^{\mu}\left(\theta,\varphi\right)}}
\end{equation}

Substituting 
\[
\Delta Pf.[r^{s}Y_{l}^{\mu}\left(\theta,\varphi\right)S_{e}\left(r\right)]=\Delta Pf.\{r^{s-l}[r{}^{\ell}\mathit{Y_{\ell}^{\mu}\left(\theta,\varphi\right)}S_{e}\left(r\right)]\}
\]

\[
\Delta Pf.[r^{s+1}\mathit{Y_{\ell}^{\mu}\left(\theta,\varphi\right)}S_{d}\left(r\right)]=\Delta Pf.\{r^{s+1-l}[r{}^{\ell}\mathit{Y_{\ell}^{\mu}\left(\theta,\varphi\right)}S_{d}\left(r\right)]\}
\]
$\Delta Pf.\Psi$ is given by the sum of the two right-hand sides,
where $\mathit{\mathit{r^{\ell}Y_{\ell}^{\mu}\left(\theta,\varphi\right)}S_{e}\left(r\right)}$
and $\mathit{r{}^{\ell}\mathit{\mathit{Y_{\ell}^{\mu}\left(\theta,\varphi\right)}}S_{d}\left(r\right)}$
are of C$^{\infty}$(R$^{3}$), so that these two Laplacians are the
products of a distribution by a function of C$^{\infty}$(R$^{3}$),
given by the usual expression of the Laplacian of the product of two
functions. It enables to show that{[}3{]}

\begin{equation}
\Delta Pf.\Psi-Pf.\Delta\Psi=\mathit{Q_{s,l}\left(\delta\right)}
\end{equation}

where

\[
\mathit{Q_{s,l}}\left(\delta\right)=\chi_{p}\left[(C_{p}+2L_{p}\ell+2L_{p}r\frac{d}{dr})S_{e}\left(r\right)\right]r{}^{\ell}Y_{\ell}^{\mu}\left(\theta,\varphi\right)\Delta^{p}\delta
\]

\[
+\chi_{q}\left[(C_{q}+2L_{q}\ell+2L_{q}r\frac{d}{dr})S_{d}\left(r\right)\right]r{}^{\ell}\mathit{Y_{\ell}^{\mu}\left(\theta,\varphi\right)}\Delta^{q}\delta
\]

\begin{equation}
p=-\frac{s+1-\ell}{2}\qquad\qquad q=-\frac{s+2-\ell}{2}
\end{equation}

These results hold in Cartesian and spherical coordinates. The Laplacian
of \textit{Pf.R}(\textit{r}) therefore reads 
\begin{equation}
(\frac{1}{r}\frac{d^{2}}{dr^{2}}r)Pf.R\left(r\right)=Pf.(\frac{1}{r}\frac{d^{2}}{dr^{2}}r)R\left(r\right))+Q_{s}\left(\delta\right)
\end{equation}

or, in terms of \textit{Pf.u}(\textit{r}) \textit{=} \textit{Pf.}$\mathit{r^{s+1}}\Sigma_{k\geq0}a_{k}r^{k}$

\begin{equation}
(\frac{1}{r}\frac{d^{2}}{dr^{2}}r)Pf.\frac{u(r)}{r}=Pf.(\frac{1}{r}\frac{d^{2}u(r)}{dr^{2}})+Q_{s}\left(\delta\right)
\end{equation}

where $Q_{s}\left(\delta\right)$ is given by (16). The Laplacian
of Pf.$\Psi$ reads

\[
\Delta Pf.\Psi=Pf.\left[(\frac{1}{r}\frac{d^{2}}{dr^{2}}r-\frac{\ell(\ell+1)}{r^{2}})\mathit{R\left(r\right)\mathit{Y_{\ell}^{\mu}\left(\theta,\varphi\right)}}\right]+Q_{s,l}\left(\delta\right)
\]

\begin{equation}
=Pf.\left[\frac{1}{r}(\frac{d^{2}}{dr^{2}}-\frac{\ell(\ell+1)}{r^{2}})\mathit{u\left(r\right)}\mathit{Y_{\ell}^{\mu}\left(\theta,\varphi\right)}\right]+Q_{s,\ell}\left(\delta\right)
\end{equation}

and the Hamiltonian of $\Psi$,

\[
HPf.\Psi=Pf.\left[(-\frac{\hbar^{2}}{2m}\frac{1}{r}\frac{d^{2}}{dr^{2}}r+\frac{\ell(\ell+1)\hbar^{2}}{2mr^{2}}+\mathit{V}_{r})R\left(r\right)\mathit{Y_{\ell}^{\mu}\left(\theta,\varphi\right)}\right]-\frac{\hbar^{2}}{2m}Q_{s,l}\left(\delta\right)
\]

\begin{equation}
=Pf.\left[\frac{1}{r}(-\frac{\hbar^{2}}{2m}\frac{d^{2}}{dr^{2}}+\frac{\ell(\ell+1)\hbar^{2}}{2mr^{2}}+\mathit{V}_{r})u\left(r\right)\mathit{Y_{\ell}^{\mu}\left(\theta,\varphi\right)}\right]-\frac{\hbar^{2}}{2m}Q_{s,\ell}\left(\delta\right)
\end{equation}

where $Q_{s,l}\left(\delta\right)$ is given by (19).

\section{The usual assumptions in R$^{3}$.}

In R$^{3}/\{0\}$ $\mathit{Q_{s}\left(\delta\right)}$ and $\mathit{Q_{s,l}\left(\delta\right)}$
vanish (and there is no symbol \textit{Pf.}), so that the equations
(4) to (9) hold. Eqs.(20) to (23) show that a necessary condition
for the usual assumptions to hold in R$^{3}$ is that $\mathit{Q_{s}\left(\delta\right)}$
= 0 = $\mathit{Q_{s,l}\left(\delta\right)}$ (it is not necessarily
the case). In order to determine $\mathit{Q_{s}\left(\delta\right)}$,
instead of (16), by substituting (12) and (14) we write $\Delta Pf.\mathit{R\left(r\right)}$in
the form

\[
\Delta Pf.R\left(r\right)=\sum_{k\geq0}a_{k}\Delta Pf.r^{k+s}=\sum_{k\geq0}a_{k}\left[Pf.\Delta r^{k+s}+\chi_{p(k)}C_{p(k)}\Delta^{p(k)}\delta\right]
\]
\[
=Pf.\Delta R\left(r\right)+\sum_{k\geq0}a_{k}\chi_{p(k)}C_{p(k)}\Delta^{p(k)}\delta\qquad\qquad p=-\frac{k+s+1}{2}
\]
which gives

\[
Q_{s}\left(\delta\right)=\sum_{k\geq0}a_{k}\chi_{p(k)}C_{p(k)}\Delta^{p(k)}\delta\qquad\qquad p=-\frac{k+s+1}{2}
\]

As $\mathit{p}\in N$, only if \textit{s} is an integer $\mathit{\leq}$
-1, and \textit{k} $\mathit{\leq}$ - \textit{s }-1, the sum amounts
to

\begin{equation}
Q_{s}\left(\delta\right)=\sum_{k=0}^{-s-1}a_{k}\chi_{p(k)}C_{p(k)}\Delta^{p(k)}\delta\qquad\qquad p=-\frac{k+s+1}{2}
\end{equation}

where all the terms of the same parity as \textit{s }are zero. As
$C_{p(k)}\neq0\neq\Delta^{p(k)}\delta$ ($p(k)\in N$ ), and as the
$\Delta^{p(k)}\left(\delta\right)$ are linearly independent, $\mathit{Q_{s}\left(\delta\right)\neq}$
0 if there is one term of the sum at least which is non zero. Hence
$\mathit{Q_{s}\left(\delta\right)}$ is non zero, and the \textit{second
assumption} on which the passage from the equations in terms of \textit{R}(\textit{r})
to the equations in terms of \textit{u}(\textit{r}) is based \textit{does
not hold} if there is at least one coefficient $a_{k}\neq0$ for which
\textit{k }+\textit{ s} is an odd negative integer, that is,

\begin{equation}
\mathit{Q_{s}\left(\delta\right)\neq}\:0\qquad\Longleftrightarrow\qquad\exists\begin{cases}
a_{k}\neq0\\
k+s=-2p-1
\end{cases}\qquad p\in N
\end{equation}

It is the case in particular if there are two non zero successive
coefficients, $a_{k}\neq0\neq a_{k+1}$, or/and if \textit{Pf.R}(\textit{r})
behaves at the origin like $\mathit{r^{s}}$, where \textit{s }is
an odd negative integer (since $\mathit{a_{o}\neq}$ 0). Thus, the
passage from the equations in terms of \textit{R}(\textit{r}) to the
equations in terms of \textit{u}(\textit{r}), in particular from (7)
to (8), which is usually based on the relation 

\begin{equation}
(\frac{1}{r}\frac{d^{2}}{dr^{2}}r)R\left(r\right)=(\frac{1}{r}\frac{d^{2}}{dr^{2}}r)\frac{u(r)}{r}=\frac{1}{r}\frac{d^{2}u(r)}{dr^{2}}
\end{equation}

does not hold if the condition (25) is satisfied. It is the case for
the solutions normalizable such that \textit{u}(0) = $\mathit{a_{o}\neq}$
0, and thus such that \textit{R}(\textit{r}) behaves at the origin
like 1/\textit{r} , since for \textit{s }= -1, $\mathit{Q_{s}\left(\delta\right)=a_{o}C_{o}\delta}$,
(21) gives

\begin{equation}
(\frac{1}{r}\frac{d^{2}}{dr^{2}}r)R\left(r\right)=(\frac{1}{r}\frac{d^{2}}{dr^{2}}r)\frac{u(r)}{r}=\frac{1}{r}\frac{d^{2}u(r)}{dr^{2}}-4\pi u(0)\delta
\end{equation}

Note that this relation, very useful, can be obtained only with the
help of the well-know relation $\Delta$(1/\textit{r}) = -4$\pi\delta$.
If \textit{s} $\in N$ indeed, except for \textit{s} = 1, \textit{$r^{s}\in$}
C$^{\infty}$(R$^{3}$), $\Delta r^{s}$ is the Laplacian in the sense
of the functions, and as $\Delta r=\Delta\left[r^{2}\left(1/r\right)\right]$,
the latter, which is the Laplacian of the product of a distribution
by a function of C$^{\infty}$(R$^{3}$), is given by the usual expression
of the Laplacian of the product of two functions, that is, $\Delta r=$
2/\textit{r}, so that if $s\in N$, in any case $\Delta r^{s}=s(s+1)r^{s-2}$.
Hence, if \textit{u}(\textit{r}) = $\Sigma_{k\geq0}a_{k}r^{k}$,

\[
\Delta\frac{u(r)}{r}=\frac{a_{o}}{r}+\Sigma_{k\geq1}a_{k}\Delta r^{k-1}=-4\pi a_{o}\delta+\Sigma_{k\geq1}a_{k}k(k-1)r^{k-3}
\]

\[
=-4\pi a_{o}\delta+\Sigma_{k\geq0}a_{k}k(k-1)r^{k-3}=\dfrac{1}{r}\frac{d^{2}u(r)}{dr^{2}}-4\pi u(0)\delta
\]

Similarly, the Laplacian of \textit{Pf.}$\Psi$ given by (17) reads

\[
\Delta Pf.\Psi={\displaystyle {\textstyle {\displaystyle {\textstyle \sum_{k\geq0}a_{k}\Delta Pf.r^{k+s}}\mathit{\mathit{Y_{\ell}^{\mu}\left(\theta,\varphi\right)}}}}}
\]

By substituting $\mathit{S_{e}\left(r\right)}=$ 1 and $S{}_{d}\left(r\right)=$
0 in (18) and (19), we have

\[
\Delta Pf.r^{s}\mathit{Y_{\ell}^{\mu}\left(\theta,\varphi\right)}=Pf.\Delta r^{s}\mathit{Y_{\ell}^{\mu}\left(\theta,\varphi\right)}\mathit{+}\chi_{p}B_{l,p}C_{p}r{}^{\ell}\mathit{\mathit{Y_{\ell}^{\mu}\left(\theta,\varphi\right)}}\Delta^{p}\delta
\]

\[
p=-\frac{s+1-\ell}{2}
\]

where $\mathit{B_{\ell,p}C_{p}=C_{p}}+2L_{p}\ell$ = {[}1- 2$\ell$/(4\textit{p}
+1){]}$\mathscr{\mathit{C_{p}}}$. As $\mathit{\ell\in N}$, $\mathit{B_{\ell,p}}$=
0 only if $\mathit{p\notin N}$, so that $\chi_{p}B_{\ell,p}C_{p}$$\neq$
0 if $p\in N$, and = 0 if $\mathit{p\notin N}$. As $\mathit{r^{\ell}\mathit{Y_{\ell}^{\mu}\left(\theta,\varphi\right)}}$
is a homogeneous polynomial of degree $\ell$, $\mathit{r^{\ell}\mathit{Y_{\ell}^{\mu}\left(\theta,\varphi\right)}}\Delta^{p}\left(\delta\right)\neq0$
only if 2\textit{p} $\geq\ell$, and then only if \textit{s}$\mathit{\leq}$
-1, so that

\[
\chi_{p}B_{l,p}C_{p}r{}^{\ell}\mathit{\mathit{Y_{\ell}^{\mu}\left(\theta,\varphi\right)}}\Delta^{p}\delta\begin{cases}
\neq0\quad & if\quad s-\ell=-2p-1\\
=0 & if\quad s-\ell\neq-2p-1
\end{cases}\qquad s\leq-1,\; p\in N
\]

In particular for \textit{s} = - ($\ell$+1), $\Delta Pf.r^{-(\ell+1}\mathit{\mathit{Y_{\ell}^{\mu}\left(\theta,\varphi\right)}\neq}$
0. Hence

\[
\Delta Pf.\Psi=\sum_{k\geq0}{\displaystyle {\textstyle {\displaystyle {\textstyle a_{k}\left[{\textstyle Pf.\Delta r^{k+s}}\mathit{Y_{\ell}^{\mu}\left(\theta,\varphi\right)}+\chi_{p(k)}B_{\ell,p(k)}C_{p(k)}\mathit{r{}^{\ell}\mathit{\mathit{Y_{\ell}^{\mu}\left(\theta,\varphi\right)}}}\Delta^{p(k)}\delta\right]}}}}
\]

\[
=Pf.\Delta\Psi+Q_{s,l}\left(\delta\right)\qquad\qquad p=-\frac{k+s+1-\ell}{2}
\]

where

\[
{\textstyle Q_{s,\ell}\left(\delta\right)=\sum_{k\geq0}a_{k}\chi_{p(k)}B_{\ell,p(k)}C_{p(k)}\mathit{r{}^{\ell}\mathit{\mathit{Y_{\ell}^{\mu}\left(\theta,\varphi\right)}}}\Delta^{p(k)}\delta}
\]

\[
p=-\frac{k+s+1-\ell}{2}
\]

with

\[
\chi_{p(k)}B_{\ell,p(k)}C_{p(k)}r{}^{\ell}\mathit{\mathit{Y_{\ell}^{\mu}\left(\theta,\varphi\right)}}\Delta^{p(k)}\delta\begin{cases}
\neq0\quad & if\quad k+s-\ell=-2p-1\\
=0 & if\quad k+s-\ell\neq-2p-1
\end{cases}
\]

where $\mathit{s\leq-}$1, $\mathit{p\in N}$. As \textit{k} $\leq$
-\textit{ s} - 1 (since 2\textit{p} $\geq\ell$), the sum amounts
to

\[
Q_{s,l}\left(\delta\right)=\sum_{k=0}^{-s-1}a_{k}\chi_{p(k)}B_{\ell,p(k)}C_{p(k)}\mathit{r{}^{\ell}\mathit{\mathit{Y_{\ell}^{\mu}\left(\theta,\varphi\right)}}}\Delta^{p(k)}\delta
\]

\begin{equation}
p=-\frac{k+s+1-\ell}{2}
\end{equation}

where the terms corresponding to even, resp. odd, values of \textit{k}
are zero , if \textit{s} and $\ell$ have the same parity, resp. different
parities. In any case, $\mathit{Q_{s,\ell}\left(\delta\right)}\neq0$
if one term of the sum at least is non zero. Hence $\mathit{Q_{s,\ell}\left(\delta\right)}$
is non zero, and the \textit{first assumption }on which the usual
resolution of the SE is based, namely, the Laplace and Hamilton operators
separate the variables so that $\Delta\Psi$ and $\mathit{H\Psi}$
amount to a radial part times $\mathit{Y_{\ell}^{\mu}\left(\theta,\varphi\right)}$
, \textit{does not hold }if there is at least one coefficient $a_{k}\neq0$
for which \textit{k} + \textit{s} -\textit{ $\ell$} is an odd negative
integer. that is,

\begin{equation}
\mathit{Q_{s,\ell}\left(\delta\right)\neq}\;0\qquad\Longleftrightarrow\qquad\exists\begin{cases}
a_{k}\neq0\\
k+s-\ell=-2p-1
\end{cases}\qquad p\in N
\end{equation}

It is the case in particular if there are two non zero successive
coefficients, $a_{k}\neq0\neq a_{k+1}$, or/and if $Pf.\Psi$ behaves
at the origin like $\mathit{r^{s}}\mathit{Y_{\ell}^{\mu}\left(\theta,\varphi\right)}$,
where \textit{s} - $\mathit{\ell}$ is an odd negative integer. Notably,
if $Pf.\Psi$ behaves at the origin like $r^{-(\mathit{\ell}+1)}\mathit{Y_{\ell}^{\mu}\left(\theta,\varphi\right)}$,
the first term is non zero,

\[
a_{o}\chi_{\ell}B_{\ell,\ell}C_{\ell}r{}^{\ell}\mathit{\mathit{Y_{\ell}^{\mu}\left(\theta,\varphi\right)}}\Delta^{\ell}\delta\neq0
\]

and hence

\begin{equation}
Q_{-(l+1),l}=\sum_{k=0}^{k=l}a_{k}\chi_{p(k)}B_{\ell,p(k)}C_{p(k)}\mathit{\mathit{r{}^{\ell}\mathit{\mathit{Y_{\ell}^{\mu}\left(\theta,\varphi\right)}}}}\Delta^{p(k)}\delta\neq0\qquad p=\frac{2\ell-k}{2}
\end{equation}

\section{The radial equations and the radial Schrödinger equation.}

One usually considers that the solutions of (7) are given by the two
roots of the indicial equation \textit{s} = $\mathit{\ell}$ and \textit{s}
= - ($\ell$+ 1), and hence that they behave at the origin, either
like $\mathit{r^{\ell}}$, or like $Pf.r{}^{-(\ell+1)}$. As they
are, either regular functions \textit{R}(\textit{r}), or singular
functions, or pseudofunctions \textit{Pf}.\textit{R}(\textit{r}),
(7) should be written

\begin{equation}
\left[-\frac{\hbar^{2}}{2m}\frac{1}{r}\frac{d^{2}}{dr^{2}}r+\frac{\ell(\ell l+1)\hbar^{2}}{2mr^{2}}+\mathit{V}_{r}\right]Pf.R\left(r\right)=\mathit{E\,}Pf.R\left(r\right)
\end{equation}

By substituting (20), the left-hand side reads

\[
\left[-\frac{\hbar^{2}}{2m}\frac{1}{r}\frac{d^{2}}{dr^{2}}r+\frac{\ell(\ell+1)\hbar^{2}}{2mr^{2}}+\mathit{V}_{r}\right]Pf.R\left(r\right)=
\]

\begin{equation}
Pf.\left[-\frac{\hbar^{2}}{2m}\frac{1}{r}\frac{d^{2}}{dr^{2}}r+\frac{\ell(\ell l+1)\hbar^{2}}{2mr^{2}}+\mathit{V}_{r}\right]R\left(r\right)-\frac{\hbar^{2}}{2m}Q_{s}(\delta)
\end{equation}

Eq.(31) can thus be satisfied only if the left-hand side involves
no expression containing $\delta$, and then if $\mathit{Q_{s}\left(\delta\right)}$=
0. Eq.(25) shows that if \textit{s }= $\ell$, as \textit{k} + $\ell$
$\neq$ -2\textit{p }-1 $\forall k$ and $\forall\ell$, $\mathit{Q_{\ell}\left(\delta\right)}$=
0 $\forall\ell$. If\textit{ s} = - ($\ell$+ 1), $Q{}_{-(\ell+1)}\neq0$
if there is at least a coefficient $\mathit{a_{k}\neq}$ 0 for which
($\ell-k$)/2 $\in N$. It is the case if there are two non zero successive
coefficients, $a_{k}\neq0\neq a_{k+1}$, or/and if $\mathit{\ell}$
is even . In this case (7) or (31) has \textit{no solutions which
behave at the origin like }$Pf.r{}^{-(\ell+1)}$). The solutions of
(7) given by the root \textit{s} = - ($\ell$+ 1) are solutions of
(7)\textit{ in R$^{3}$\{/0\}, but not necessarily in R$^{3}$}.

The radial equation which has solutions which behave at the origin
like $\mathit{r^{\ell}}$ or $Pf.r{}^{-(\ell+1)}$, is the \textit{extension
}in R$^{3}$ of (7), that is, the equation 

\begin{equation}
Pf.\left[-\frac{\hbar^{2}}{2m}\frac{1}{r}\frac{d^{2}}{dr^{2}}r+\frac{\ell(\ell+1)\hbar^{2}}{2mr^{2}}+\mathit{V}_{r}\right]R\left(r\right)=\mathit{E\,}Pf.R\left(r\right)
\end{equation}

and similarly, the radial equation which has solutions which behave
at the origin like $r{}^{\ell+1}$ or $Pf.r{}^{-\ell}$, is the extension\textit{
}in R$^{3}$ of (8), that is, the equation

\begin{equation}
Pf.\left[-\frac{\hbar^{2}}{2m}\frac{d^{2}}{dr^{2}}+\frac{\ell(\ell+1)\hbar^{2}}{2mr^{2}}+\mathit{V}_{r}\right]u\left(r\right)=\mathit{E}\, Pf.\mathit{u}(\mathit{r})
\end{equation}

(In other words, \textit{Pf.R}(\textit{r}) is a solution of (33),
and \textit{Pf.u}(\textit{r}) a solution of (34), if in R$^{3}$/\{0\}
\textit{R}(\textit{r}) is a solution of (7), and \textit{u}(\textit{r})
a solution of (8), so that the solutions of (33) and (34) are given
by the two roots \textit{s}$=\ell$ and \textit{s} = - ($\ell$+1)).
By substituting (33) in (23) we obtain

\[
HPf.\Psi=Pf.\left[-\frac{\hbar^{2}}{2m}\frac{1}{r}\frac{d^{2}}{dr^{2}}r+\frac{\ell(\ell+1)\hbar^{2}}{2mr^{2}}+\mathit{V}_{r}\right]R\left(r\right)\mathit{Y_{\ell}^{\mu}\left(\theta,\varphi\right)}-\frac{\hbar^{2}}{2m}Q_{s,l}\left(\delta\right)
\]

\begin{equation}
=E\: Pf.\left[R\left(r\right)\mathit{Y_{\ell}^{\mu}\left(\theta,\varphi\right)}\right]-\frac{\hbar^{2}}{2m}Q_{s,\ell}\left(\delta\right)
\end{equation}

and by substituting (34) in (23),

\[
HPf.\Psi=Pf.\left[\frac{1}{r}(-\frac{\hbar^{2}}{2m}\frac{d^{2}}{dr^{2}}+\frac{\ell(\ell+1)\hbar^{2}}{2mr^{2}}+\mathit{V}_{r})u\left(r\right)\mathit{Y_{\ell}^{\mu}\left(\theta,\varphi\right)}\right]-\frac{\hbar^{2}}{2m}Q_{s,l}\left(\delta\right)
\]

\begin{equation}
=E\: Pf.\left[\frac{u(r)}{r}\mathit{Y_{\ell}^{\mu}\left(\theta,\varphi\right)}\right]-\frac{\hbar^{2}}{2m}Q_{s,\ell}\left(\delta\right)
\end{equation}

It shows that depending on whether or not $\mathit{Q_{s,\ell}\left(\delta\right)}$
is zero, and hence depending on whether or not the condition (29)
is satisfied, the radial equations are \textit{derived from two different
equations}.

We can therefore conclude. When the solutions of the radial equations
are given by the root \textit{s} =$\ell$, they are regular functions,
the symbol $\mathit{Pf.}$ is useless, as \textit{k} $\mathit{\neq}$
-2\textit{p} -1 $\forall k$, $\mathit{Q_{\ell,\ell}\left(\delta\right)}$=
0 $\forall\ell$, the radial equations are derived from the SE, the
equations (4) to (9) hold in R$^{3}$.

When the solutions are given by the root \textit{s} = - ($\ell$+1),
they are singular functions, or pseudofunctions, and as $Q_{-(l+1),l}\mathit{\left(\delta\right)}\neq$
0, eq.(30),\textit{ the radial equations are not derived from the
SE, but from the equation}

\begin{equation}
HPf.\Psi=E\, Pf.\Psi-\frac{\hbar^{2}}{2m}Q_{-(\ell+1),\ell}\left(\delta\right)
\end{equation}

In this case \textit{the radial equations are not the radial parts
of the SE}, this is why their solutions substituted in $Pf.\Psi$
do not give solutions of the SE, but solutions of (37). For $\ell$
> 0 these solutions are not normalizable, and then generally not taken
into account, so that the fact that the radial equations (7) and (8),
that is, (33) and (34), are not derived from the SE, and then do not
hold, is of no consequence.

The problem occurs in the radial case owing to the fact that the solutions
are normalizable, and then must be taken into account. As the symbol
$\mathit{Pf.}$ can be dropped for $\Psi$ and \textit{R}(\textit{r})
which behave at the origin like 1/\textit{r} , but not for the operator
$\mathit{Pf.}\Delta$, (33) reads

\begin{equation}
Pf.\left[-\frac{\hbar^{2}}{2m}\frac{1}{r}\frac{d^{2}}{dr^{2}}r+\mathit{V}_{r}\right]R\left(r\right)=\mathit{E}\, R\left(r\right)
\end{equation}

It is derived from the equation (35) for $\ell$= 0, that is,

\[
H\Psi=\frac{1}{\sqrt{4\pi}}\: Pf.\left[-\frac{\hbar^{2}}{2m}\frac{1}{r}\frac{d^{2}}{dr^{2}}r+\mathit{V}_{r}\right]R\left(r\right)+\frac{\hbar^{2}\sqrt{\pi}}{m}u(0)\delta
\]

\begin{equation}
=\frac{1}{\sqrt{4\pi}}\: E\: R\left(r\right)+\frac{\hbar^{2}\sqrt{\pi}}{m}u(0)\delta
\end{equation}

As $\mathit{u}\left(0\right)=a_{o}\neq0,$ \textit{u}(\textit{r})
is a regular function, \textit{Pf.}(\textit{d\texttwosuperior{}u}(\textit{r})\textit{/dr\texttwosuperior{}})$\equiv$(\textit{d\texttwosuperior{}u}(\textit{r})\textit{/dr\texttwosuperior{}}),
(34) is written in the form (9). The latter is derived from (36) for
$\ell$= 0, which reads, in agreement with (27),

\[
H\Psi=\frac{1}{\sqrt{4\pi}}\:\frac{1}{r}\left[-\frac{\hbar^{2}}{2m}\frac{d^{2}}{dr^{2}}r+\mathit{V}_{r}\right]u\left(r\right)+\frac{\hbar^{2}\sqrt{\pi}}{m}u(0)\delta
\]

\begin{equation}
=\frac{1}{\sqrt{4\pi}}\: E\:\frac{u(r)}{r}+\frac{\hbar^{2}\sqrt{\pi}}{m}u(0)\delta
\end{equation}

Thus, usually (9) is considered as derived from the SE, and hence
as the reduced SE, and the question is to justify that its solutions
substituted in $\Psi$ give solutions of the SE, and are admissible,
when \textit{u}(0) = 0, but not when\textit{ u}(0)\textit{ $\neq$}
0. Eq.(40) shows that in fact (9) is derived from the SE and is the
reduced SE when \textit{u}(0) = 0, but that\textit{ when u(0) $\neq$0
(9) is not the reduced SE, it is not derived indeed from the SE, but
from the equation} 

\begin{equation}
H\Psi=E\:\Psi+\frac{\hbar^{2}\sqrt{\pi}}{m}u(0)\delta
\end{equation}

This is why the solutions of the radial equation (9) substituted in
$\Psi$ give solutions of the SE when\textit{ u}(0) = 0 , but not
when \textit{u}(0) = $\mathit{a_{o}\neq}$ 0. In this case they give
solutions of (41).

Eqs.(20) to (23) show that the usual misunderstandings occur when
the Laplacian of $\Psi$ in (4) and (5), and hence in the SE (3),
and the Laplacian of \textit{R}(\textit{r}) in (7), are not taken
in the sense of the distributions, but of the functions. The radial
equations (7) and (8) are then considered as derived from the SE \textit{regardless
of their solutions}. It can be summarized as follows : the solutions
of the radial SE satisfy the condition \textit{u}(0) = 0, they are
normalizable, and they are solutions of the radial equation (9), but
the normalizable solutions of (9) such that \textit{u}(0) $\neq$
0 are not solutions of the radial SE. This is why \textit{the boundary
condition u(0) = 0 is aimless.}

\section{Concluding remarks}

The radial equations are usually written, either in terms of\textit{
R}(\textit{r}), or in terms of \textit{u}(\textit{r}), the two equations
being considered as equivalent. In fact it is not the case, and not
only because the latter is derived from the former. Eq.(32) shows
that when $\mathit{Q_{-(\ell+1)}(\delta)\neq}$ 0, the solutions of
(33) which behave at the origin like\textit{ }$Pf.r{}^{-(\ell+1)}$
are not solutions of (31), but of the equation

\[
\left[-\frac{\hbar^{2}}{2m}\frac{1}{r}\frac{d^{2}}{dr^{2}}r+\frac{\ell(\ell+1)\hbar^{2}}{2mr^{2}}+\mathit{V}_{r}\right]Pf.R\left(r\right)=
\]

\begin{equation}
\mathit{E\,}Pf.R\left(r\right)-\frac{\hbar^{2}}{2m}Q_{-(\ell+1)}(\delta)
\end{equation}

As the radial equations (33) and (34) are equivalent, to avoid any
confusion between (7) or (31) and (33), it may be preferable to write
the radial equations in terms of $u\left(r\right)$, all the more
as in the case of the normalizable solutions these equations are written
without the symbol \textit{Pf.}. Besides, if \textit{u}(\textit{r})
is a function singular at the origin, or a pseudofunction \textit{Pf.u}(\textit{r}),
the term (\textit{d}\texttwosuperior{}/\textit{dr}\texttwosuperior{})\textit{Pf.u}(\textit{r})
is not defined in the theory of distributions (and \textit{a fortiori}
of the functions), so that (9) holds in R$^{3}$ only for regular
functions, which is just the case of normalizable solutions to which
we confine ourselves.

What we called the ``usual resolution of the Schrödinger equation''
is the process{[}4{]} in which the solutions of this equation are
obtained by substituting in $\Psi$, either the solutions \textit{R}(\textit{r})
of (7), or the solutions \textit{u}(\textit{r}) of (8), where (7)
has been derived from the SE (3), and (8) from (7), by taking the
Laplacians of $\Psi$ and of \textit{R}(\textit{r}) in the sense of
the \textit{functions}, while considering that there are \textit{singular}
\textit{solutions} which must be ruled out. In other words, in no
handbook of quantum mechanics, (nor article to my knowledge) the radial
equation (7) is derived from the SE (3), and the radial equation (8)
from (7), by taking the Laplacians of $\Psi$ and of \textit{R}(\textit{r})
in the sense of the distributions, as it is required. Singular solutions
are sometimes ruled out by considering that if \textit{R}(\textit{r})
is a singular solution of (7), $\Psi$ cannot be a solution of the
SE since $\Delta\Psi$ involves an expression containing $\delta$.
But if $\Delta\Psi$ involves an expression containing $\delta$,
it shows that (7) which does not involve itself any expression containing
$\delta$, cannot be derived from the SE, in this case indeed\textit{
R}(\textit{r}) is not a solution of (7), as it is assumed, but of
(42). 

It is thus because the Laplacians are not taken in the sense of the
distributions that the two usual assumptions on which this resolution
is based, and then the equations that they imply, hold in R$^{3}$/\{0\},
but not necessarily in R$^{3}$. It is what constrains to impose the
supplementary boundary condition \textit{u}(0) = 0. Justifying this
condition has always been a requirement (See e.g.{[}5{]}), and remains
a present-day problem (See e.g.{[}6{]}). The various and often conflicting
arguments proposed to justify this condition are based on this misunderstanding,
they never led to a consensus. Insofar as they are a part of the usual
resolution of the SE, these arguments will be also revisited.

\part*{Acknowledgments}

I am grateful to A. Khelif for clarifications regarding some mathematical
questions.

\end{document}